\documentclass[10pt,twocolumn]{article}

\usepackage{scicite}
\usepackage[pdftex]{graphicx}
\usepackage{amssymb,amsmath}

\usepackage{times}

\newenvironment{sciabstract}{%
\begin{quote} \bf}
{\end{quote}}

\newcounter{lastnote}
\newenvironment{scilastnote}{%
\setcounter{lastnote}{\value{enumiv}}%
\addtocounter{lastnote}{+1}%
\begin{list}%
{\arabic{lastnote}.}
{\setlength{\leftmargin}{.22in}}
{\setlength{\labelsep}{.5em}}}
{\end{list}}

\title{Deterministic Preparation of a Tunable
Few-Fermion System}
\author{F.\,Serwane$^{1,2,3\,*\,\dagger}$, G. Z\"urn$^{1,2\,\dagger}$, T. Lompe$^{1,2,3}$,T.\,B. Ottenstein$^{1,2,3}$,\\ A.\,N. Wenz$^{1,2}$ \& S. Jochim$^{1,2,3}$
\\
\\
\small{$^{1}$Physikalisches Institut, Ruprecht-Karls-Universit\"at, 69120 Heidelberg, Germany}\\
\small{$^{2}$Max-Planck-Institut f\"ur Kernphysik, Saupfercheckweg 1, 69117 Heidelberg, Germany}\\
\small{$^{3}$ExtreMe Matter Institute EMMI, GSI Helmholtzzentrum f\"ur Schwerionenforschung, 64291 Darmstadt, Germany}\\
\\
\small{$^\ast$To whom correspondence should be addressed; E-mail:  friedhelm.serwane@mpi-hd.mpg.de.}\\
\small{$\dagger$These authors contributed equally to this work.}
}

\date{}

\begin{document} 


%\baselineskip12pt

\twocolumn[
  \begin{@twocolumnfalse}
    \maketitle
    \begin{sciabstract}
    Systems consisting of few interacting fermions are the building blocks of matter, with atoms and
nuclei being the most prominent examples. We have created a few-body quantum system with
complete control over its quantum state using ultracold fermionic atoms in an optical dipole trap.
Ground-state systems consisting of 1 to 10 particles were prepared with fidelities of $\sim$ 90\%. We
can tune the interparticle interactions to arbitrary values using a Feshbach resonance and observed
the interaction-induced energy shift for a pair of repulsively interacting atoms. This work is
expected to enable quantum simulation of strongly correlated few-body systems.      
    \end{sciabstract}
  \end{@twocolumnfalse}
  ]
The exploration of naturally occurring few-body quantum systems such as atoms and nuclei has been extremely successful, largely because they could be prepared in well defined quantum states. \begin{figure}[htb]
\centering
\includegraphics[width=79mm]{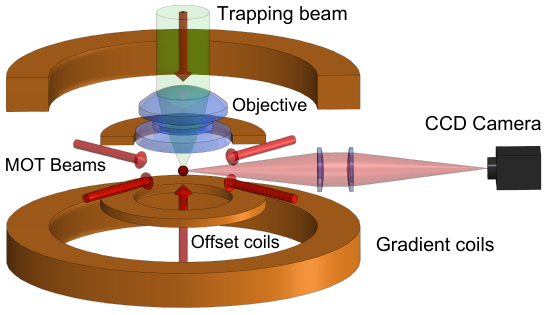}
\caption{Experimental setup. Systems with up to 10 fermions are prepared with $^{6}$Li atoms in a micrometer-sized optical dipole trap created by the focus of a single laser beam.
The number of atoms in the samples is detected with single atom resolution by transferring them into a compressed magneto-optical-trap (MOT) and collecting their fluorescence on a CCD camera. A Feshbach resonance allows one to tune the interaction between the particles with a magnetic offset field.
}
\end{figure}Because these systems have limited tunability, researchers created quantum dots---``artificial atoms''---in which properties such as particle number, interaction strength and confining potential can be tuned\cite{Heer1993,Reimann2002}. However, quantum dots are generally strongly coupled to their environment which hindered the deterministic preparation of well-defined quantum states. \\In contrast, ultracold gases provide tunable systems in a highly isolated environment\cite{Bloch2008,Ketterle2008}.
They have been proposed as a tool for quantum simulation\cite{Lloyd1996, Jaksch1998} which has been realized experimentally for various many-body systems\cite{Greiner2002,Jordens2008,Schneider2008,Nascimbene2010}.
Achieving quantum simulation of few-body systems is more challenging because it requires complete control over all degrees of freedom: the particle number, the internal and motional states of the particles, and the strength of the inter-particle interactions.
One possible approach to this goal is using a Mott insulator state of atoms in an optical lattice as a starting point. In this way, systems with up to four bosons per lattice site have been prepared in their ground state\cite{Sherson2010,Bakr2010}. 
Recently, single lattice sites have been addressed individually\cite{Weitenberg2011}. 
In single isolated trapping geometries, researchers could suppress atom number fluctuations by loading bosonic atoms into small-volume optical dipole traps\cite{frese2000,schlosser2001,chuu2005,Weber2006,Grunzweig2010}. 
However, these experiments were not able to gain control over the system's quantum state.\\
We prepare few-body systems consisting of 1 to 10 fermionic atoms in a well-defined quantum state making use of Pauli's principle, which states that each single-particle state cannot be occupied by more than one identical fermion. Therefore, the occupation probability of the lowest energy states approaches unity for a degenerate Fermi gas, and we can control the number of particles by controlling the number of available single-particle states. We realize this by deforming the confining potential such that quantum states above a well defined energy become unbound.
This approach requires a highly degenerate Fermi gas in a trap whose depth can be controlled with a precision much higher than the separation of its energy levels.\\
To fulfill these requirements, we use a small volume optical dipole trap with large level spacing. This microtrap is created by the focus of a single laser beam (fig. 1) with a waist of $w_0\lesssim1.8\,\mu$m and measured radial and axial trapping frequencies $(\omega_r,\omega_a) =2 \pi \times (14.0\pm0.1,1.487\pm0.010)$\,kHz\cite{SOM}.
We load the microtrap from a reservoir of cold atoms. 
The reservoir consists of a two-component mixture of $^{6}$Li atoms in the two lowest energy Zeeman substates $\vert F=1/2,m_F=+1/2\rangle$ and $\vert F=1/2,m_F=-1/2\rangle$ (labeled state $\vert1\rangle$ and $\vert2\rangle$) in a large volume optical dipole trap. The reservoir has a degeneracy of $T/T_F\approx0.5$\cite{SOM}, where $T_F$ is the Fermi temperature.
We superimpose the microtrap with the reservoir and transfer about 600 atoms into the microtrap. After removal of the reservoir, the degeneracy of the system is determined by $T_F\approx3\,\mu$K in the microtrap and the temperature $T\lesssim250$\,nK of the reservoir\cite{viverit2001}. Assuming thermal equilibrium between the microtrap and the reservoir this corresponds to $T/T_F\lesssim0.08$. According to Fermi-Dirac statistics this yields an occupation probability for the lowest state exceeding $0.9999$ which is large enough not to constrain our preparation scheme.\\   
To spill the excess atoms from the microtrap we add a linear potential in the axial direction by applying a magnetic field gradient. To obtain the same potential for both components we apply the gradient at a large magnetic offset field where the difference in the magnetic moments of states $\vert 1 \rangle$ and $\vert2\rangle$ is negligible. A particular magnetic field is then chosen so that the interaction strength between atoms in the different states vanishes because of a nearby Feshbach resonance (fig. S2). By varying the depth of the microtrap and the strength of the magnetic field gradient we can control the number of bound states in the potential (fig. 2A). If fewer than 10 bound states remain, the system is essentially one-dimensional because of the approximate $1$:$10$ aspect ratio of the trap; consequently, each energy level is occupied by one atom per spin state. During the spilling process we adiabatically tilt the potential, wait to let atoms in unbound states escape and then ramp the potential back up\cite{SOM}.
\\To probe the prepared systems, it is necessary to measure the number of atoms in the microtrap with single atom resolution and near unity fidelity. We achieve this by releasing the atoms from the microtrap, recapturing them in a compressed magneto-optical trap and then recording their fluorescence with a charge-coupled device (CCD) camera (fig. 1).
With this technique we can count the total number of atoms in the magneto-optical trap with a fidelity exceeding $99\%$ for 1 to 10 atoms (fig. S1).\\
Figure 2 shows the mean atom number and its variance as a function of the minimum microtrap depth during the spilling process. The atom number shows a step-like dependence on the trap depth with plateaus for even atom numbers. These plateaus appear at trap depths where the potential barrier for atoms in the uppermost level becomes so low that these atoms leave the trap on a timescale that is shorter than the duration of the spilling process. A simple estimation\cite{ottenstein2010} shows that the lifetime of this state can be up to three orders of magnitude shorter than the lifetime of the lower states.  
When an appropriate trap depth is chosen the fluctuations in the atom number are as low as $var/\langle N\rangle=\sigma^2/\langle N\rangle = 0.017$ for eight atoms, corresponding to a suppression of 18\,dB compared to a system obeying Poissonian statistics. We can then calculate an upper bound for the degree of degeneracy in the microtrap of $T/T_F<0.19$ by assuming that all fluctuations result from holes in the Fermi distribution; this provides a complementary method to probe the degeneracy of the lowest energy states of an ultracold Fermi gas which is conceptually related to recent studies of antibunching in degenerate Fermi gases\cite{Sanner2010,Mueller2010}. \\
\begin{figure}[h!]
\centering
\includegraphics[width=79mm]{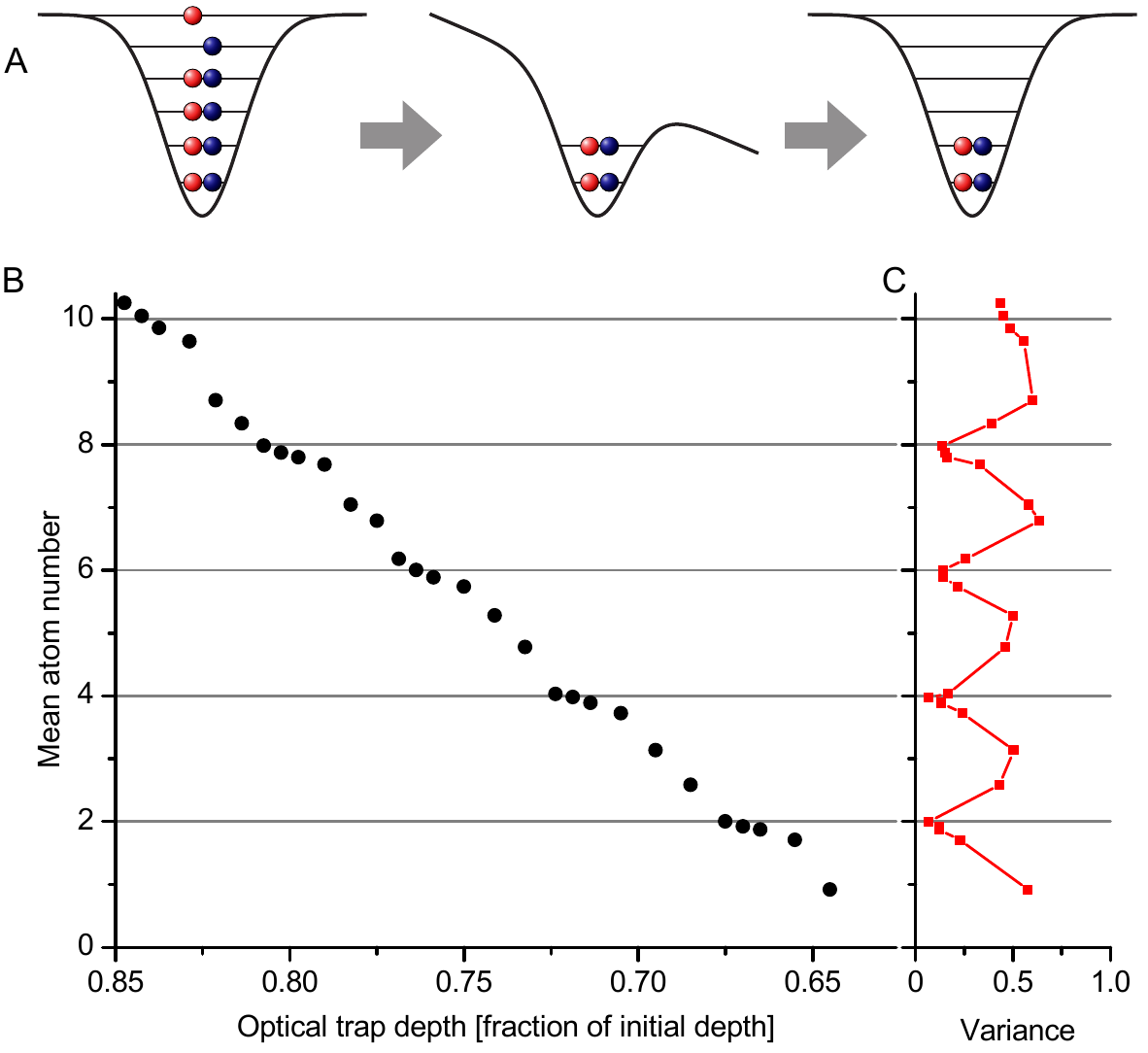}
\caption{(A) The spilling process. Starting from a degenerate two-component Fermi gas of about 600 atoms in the microtrap, we create few-particle samples by adiabatically deforming the potential to spill atoms in higher levels. After the potential has been restored, the system is in a well defined few-particle quantum state.
(B) Controlling the number of quantum states. When the trap depth is reduced, the mean atom number decreases in steps of two because each energy level in the trap is occupied with one atom per spin state. Each data point is the average of $\sim$ 190 measurements with $\sigma$ as the standard deviation and $var=\sigma^2$ as the variance (shown on the right) (C). For even atom numbers, the number fluctuations are strongly suppressed. For eight atoms, we achieve a suppression of 18\,dB of $var/\langle N\rangle$ compared to a system obeying the Poissonian statistics.}
\end{figure}
To estimate the probability of finding the system in its ground state after the spilling process we bin the measured fluorescence signal into a histogram (fig. 3B). For the preparation of systems consisting of two fermions we obtain a fidelity of $96$($1$)\%. The error is the statistical error calculated by assuming that the occurrence of samples with undesired atom number follows a Poissonian distribution. From combinatorial considerations\cite{SOM} we deduce that only a negligible fraction of the prepared two-particle systems are not in the ground state before we ramp the potential back up at the end of the preparation process. To check whether we create excitations in the system by ramping up the potential, we perform the spilling process a second time. After the second spilling process we measure a fidelity of $92$($2$)\% for preparing two atoms. This yields an upper bound of $6$($2$)\% for the excitation probability during the potential ramps\cite{SOM}. If we assume the same excitation probability for ramping up and down we get an estimated  fidelity of $93$($2$)\% to prepare the system in its ground state after ramping the potential back up after the first spilling process. For eight atoms we find a ground state preparation probability of $84$($2$)\%.
By varying the time between the two spilling processes we found the $1/e$-lifetime of the prepared two-particle system in its ground state to be $\sim60$\,s which shows the high degree of isolation from the environment.\\  
\begin{figure}[h!]
\centering
\includegraphics[width=79mm]{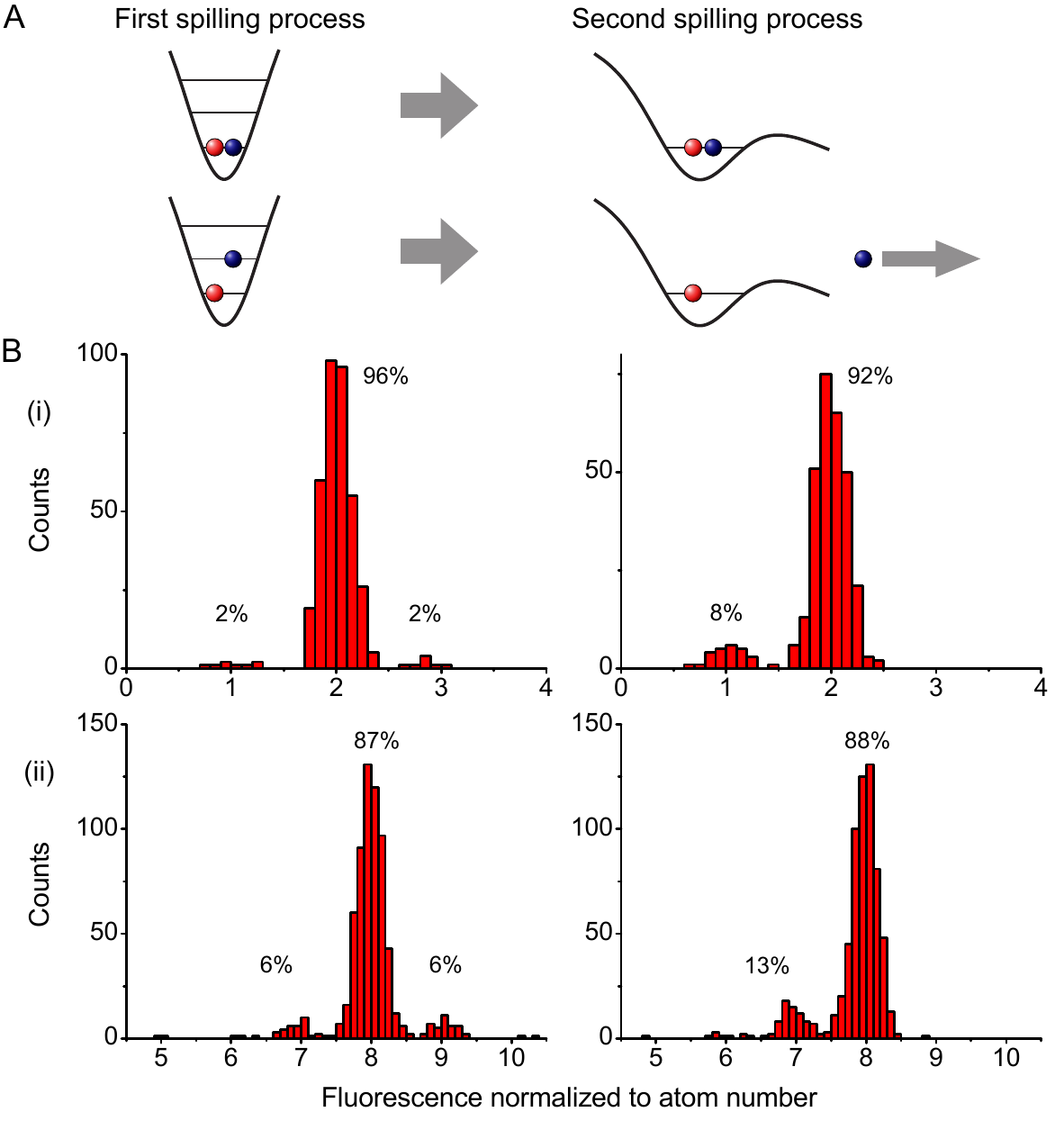}
\caption{(A) Fidelity of preparing systems in the ground state.
To determine how many of the prepared few particle systems are in their ground state, we repeat the spilling process. This removes atoms in higher levels but leaves the ground state unchanged.
(B) Histograms after the first and second spilling process for the preparation of (i) two atoms and (ii) eight  atoms. The numbers above the peaks give the relative occurrences of the counts within the corresponding peaks. The fidelity after the second spilling process (right) remains almost unchanged, indicating that the ground state is prepared with high fidelity.}
\end{figure}
To realize configurations with an arbitrary imbalance in the number of atoms in state $\vert 1 \rangle$ and $\vert 2 \rangle$ we prepare balanced systems and perform a second spilling process that only removes atoms in state $\vert1\rangle$. We do this by changing the value of the magnetic offset field to $40$ G where atoms in state $\vert2\rangle$ have negligible magnetic moment and are therefore unaffected by the magnetic field gradient (fig. S2). Using this technique we have created imbalanced systems with fidelities similar to those in the balanced case \cite{SOM}.\\
Precise control over the trapping potential is not only essential to prepare few-body samples, it is also an effective tool to probe strongly interacting systems.
We use it to explore one of the simplest non-trivial few-body systems: two non-identical fermions with repulsive interactions in the ground state of a one-dimensional harmonic trap. 
We first prepare a non-interacting pair of atoms in states $\vert 1 \rangle$ and $\vert 2 \rangle$ in the ground state of the trap.
Then we lower the potential barrier such that the two atoms slowly escape the trap on a timescale of $\tau=630\pm120$\,ms which we measure by recording the decrease in the mean atom number as a function of hold time (fig. 4B).\\
\begin{figure}[h!]
\centering
\includegraphics[width=70mm]{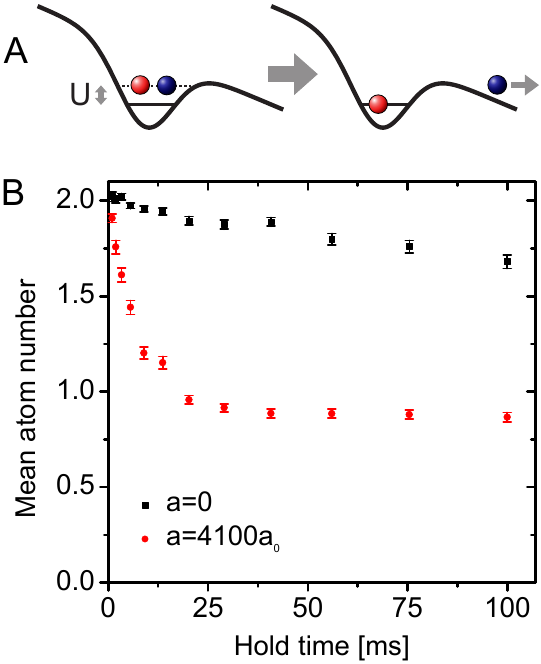}
\caption{(A) By tuning the barrier height, we can prepare systems where the atoms leave the trap with a well defined rate. Interactions cause an energy shift $U$ and thereby change the effective height of the potential barrier. 
(B) The barrier is chosen such that two non-interacting atoms ($a=0$) escape from the trap on a timescale of $\tau=630\pm120$\,ms (black squares). With repulsive interactions ($a=4100\,a_0$) we observe a much faster escape of one atom. After this atom has left the trap, the loss of the second atom becomes suppressed (red circles). Each data point is the average of $\sim$ $190$ measurements. The error bars show the standard error of the mean.}
\end{figure}
To repeat the measurement for two atoms with repulsive interaction, we tune the scattering length $a$ to a large positive value, $a=4100\,a_0$, where $a_0$ is the Bohr radius, using a Feshbach resonance. In this case we observe a much higher initial loss rate followed by a slow decay.
This fast initial loss can be explained by the energy shift $U$ of the ground state due to repulsive interactions which effectively decreases the height of the potential barrier (fig. 4A). Because the system is quasi-one-dimensional for the lowest 10 axial states, one has to consider the radial confinement\cite{olshanii1998} for the calculation of $U$\cite{Busch1998}. Given the trap parameters one expects a shift $U$ on the order of half the axial level spacing per particle. After one of the atoms has left the trap, the interaction energy drops to zero, leaving the remaining atom in the unperturbed ground state of the potential. Within our measurement accuracy, we measure an equal probability for the remaining atom to be in state $\vert 1 \rangle$ and $\vert2\rangle$. By developing theoretical models for these interaction-induced dynamics one can use this method to quantitatively study strongly interacting systems.\\
The system we created is well suited for quantum simulation with fully controlled few-body systems. For attractive interactions, it can be used to study BCS-like pairing in finite systems which is a model used for the description of nuclei\cite{migdal1959}.
Splitting a trap containing a repulsively interacting pair of atoms into a double well creates entangled pairs of neutral atoms, which can be used for quantum information processing\cite{hayes2007,raizen2009}.
By transferring the prepared ground-state samples into a periodic potential\cite{Zimmermann2011}, our system can be used to overcome one of the current major challenges of studying quantum many-body physics with ultracold atoms: preparing systems with sufficiently low entropy to explore phenomena such as quantum magnetism.

\bibliography{few_fermions}

\bibliographystyle{Science}

% Following is a new environment, {scilastnote}, that's defined in the
% preamble and that allows authors to add a reference at the end of the
% list that's not signaled in the text; such references are used in
% *Science* for acknowledgments of funding, help, etc.

\begin{scilastnote}
\item
\textbf{Acknowledgments}.\\
We thank M.\,G. Raizen for inspiring discussions and M. Weidem\"uller for the loan of a fibre laser. This work was supported by the Helmholtz Alliance HA216/EMMI and the Heidelberg Center for Quantum Dynamics.
We thank J. Ullrich and his group for their support. G.Z. and A.N.W. acknowledge
support by the IMPRS-QD.
\end{scilastnote}

\clearpage

\section*{Methods}
\subsection*{Preparation of the few-fermion samples.}

We prepare a reservoir of cold atoms by evaporative cooling of a spin mixture of states $\vert 1 \rangle$  and $\vert 2 \rangle$ at a magnetic field of $B = 300$\,G as described in previous work\cite{Ottenstein2008}. The reservoir consists of approximately $2\cdot10^4$ atoms per spin state in a crossed beam optical dipole trap ($\lambda = 1070$\,nm, waist $w_0 = 40\,\mu$m, crossing angle $15^\circ$) with trapping frequencies $\omega_r \approx 2\pi\times 370$\,Hz and $\omega_l \approx 2\pi\times 34$\,Hz. According to a Gaussian fit to the in-situ density distribution the temperature in the reservoir is $T\lesssim250$\,nK, which corresponds to $T/T_F \approx 0.5$.
To load the microtrap we turn it on within $100$\,ms and wait for $20$ ms to let the sample thermalize. Then we ramp the magnetic field to $B=523$\,G, where the scattering length crosses zero to switch off the inter-particle interaction and subsequently remove the reservoir.\\
\subsection*{Properties of the microtrap.}
The microtrap is created by the focus of a single laser beam ($\lambda = 1064$\,nm). It is focused with a two-lens objective with a numerical aperture of $0.36$. The beam waist in the focus is calculated from the measured trap frequencies to be $w_0\approx1.8\,\mu$m. As the trap contains only few radial levels, the actual potential deviates from the harmonic approximation. To determine the energy levels in the trap for the lowest levels we prepare systems of two non-interacting atoms in the lowest axial level of the trap. Then we modulate the microtrap position by shaking a mirror in the optical path of the microtrap setup with frequency $\omega$ for a defined period. If $\omega$ equals to the difference in frequency of the first and second level $\Delta E_{12}/\hbar$, atoms are excited to the second level.
We find $\Delta E_{12,axial}/\hbar=2 \pi \times (1,487\pm0,010)\,$kHz and $\Delta E_{12,radial}/\hbar=2 \pi \times (14,0\pm0,1)\,$kHz.
\subsection*{Spilling scheme.}
To spill the atoms from the microtrap we first ramp up a magnetic field gradient of $B'~=~18.9(2)~\,$ Gauss/cm within $150$ ms such that approximately the ten lowest states remain bound in the trap. The maximum slope of the gradient is smaller than $1.5\,\text{Gauss}/\text{cm}/\text{ms}$ which is slow compared to the frap frequency and therefore the sample is not heated during the ramp. The fine tuning of the potential is done $20\,$ms after the gradient has reached its final value by changing the depth of the optical trap. The spilling process consists of a linear ramp over $8$\,ms of the optical potential starting from 100\% of the trap depth, corresponding to a laser power of P$= 265(27)\, \mu$W, to 63-84$\%$ of the original depth (fig. 2B), waiting for a hold time of $25$\,ms to let the atoms escape from the trap and linearly ramping the potential back to its original value within $8$ ms.
The potential of the tilted microtrap in the z-direction can be written as 
\begin{eqnarray*}
V(z)=V_{opt}(z)+V_{mag}(z)\\
=V_0\,(1-\frac{1}{(1 + (z/z_r)^2))})-B'\mu z
\end{eqnarray*}
 where $V_0/k_b\approx3\,\mu$K is the maximum depth at the center of the optical trap, $z_r=\pi\, w_0^2/\lambda$ is the Rayleigh range and $\mu$ is the magnetic moment of the atoms.

\subsection*{High fidelity atom number detection.}
We detect the number of atoms in the prepared samples by recapturing them in a magneto-optical trap (MOT) and collecting their fluorescence signal\cite{Hu1994}. The magnetic field gradient has a strength of $250$\,G/cm, the diameter of the MOT beams is $\sim 4$\,mm and their frequency is red detuned from the resonance by twice the natural linewidth of the transition. While we cannot determine the recapture efficiency of the MOT directly, it cannot be lower than the highest measured preparation fidelity per atom of $98(1)$\%. To record the fluorescence from the MOT we image it onto a CCD camera with an imaging system with numerical aperture of $0.17$, capturing about $1$\% of the emitted photons. During the $0.5$\,s exposure time of the CCD one atom scatters about $1.9\times10^6$ photons. Considering the numerical aperture and the quantum efficiency of the imaging system roughly $1\times10^4$ photons per atom are detected. The exposure time is much shorter than the $1/e$-lifetime  of $250$\,s of the atoms in the MOT measured for 8 atoms. This lifetime is long enough that neither light-induced collisions in the MOT nor collisions with background gas atoms limit our detection fidelity. To deduce the atom number from the fluorescence signal we bin all data from each series of measurements into one histogram. These histograms show distinct peaks, each corresponding to an integer number of atoms in the MOT. From the spacing of the peaks we extract the calibration factor for the mean fluorescence per atom. Because of fluctuations of the intensity of the MOT beams and the detuning of the MOT the fluorescence signal drifts on a few percent level on a timescale of several minutes. To compensate for this drift, we rescale the fluorescence signal of each measurement by a factor which is obtained by taking the average fluorescence per atom of the ten previous and following measurements. To obtain this average for rescaling we only consider data with fluorescence signals that are close to a peak in the histogram, i.e. maximum $1\sigma$ distance. Then, the rescaled atom numbers are binned into a histogram (fig. S1) and Gaussians are fitted to the peaks. We find a separation of the peak centers of $\sim 6 \sigma$. The data points within 2$\sigma$ of a peak center are binned to integer values which represent the number of atoms in the prepared sample. The $5$\% of measurements outside of the two-sigma width of the peaks are rejected. This is possible because the measurements are uncorrelated and the atom number detection is independent from the preparation. 

\subsection*{Estimating the ground state preparation fidelity.}
To estimate the fidelity for preparing two atoms in the ground state of the trap from the histogram shown in figure 4B we only consider the lowest levels of the trap. This is based on the assumption that atoms in higher levels have only negligible probability to remain trapped after the spilling process.\\
From the number of prepared samples containing one or three atoms we can deduce upper bounds for the probability to find an atom missing in the lowest level or an atom remaining in the second level of $2$\% each. As the atoms are non-interacting during the preparation process we can assume these probabilities to apply to each atom individually. A system containing two atoms which are not in the ground state requires both a hole in the lowest level and an atom on the second level. This is suppressed by a factor of ($0.02$)$^2=4\cdot 10^{-4}$. From this we conclude that only a negligible fraction of the observed two-atom samples were not prepared in their ground state.\\
However this does not exclude the possibility of exciting the system when closing the trap at the end of the spilling process. Therefore we perform the spilling process a second time, which removes atoms in higher levels of the trap. We find that the second spilling process reduces the number of samples containing two atoms from $96$($1$)\% to $92$($2$)\%. If we assume that the $2$\% probability of preparing one atom after the first spilling process is due to states beeing non-occupied before the spilling process and account for the fact that almost all samples containing three atoms will have two atoms after the second spilling process, we would expect $98$($2$)\% of the samples to contain two and $2$($2$)\% of the samples to contain one atom. This leads us to the conclusion that there is a $6$($2$)\% probability to create an excitation while ramping the barrier up and back down.\\
Following the same consideration for a system of eight atoms we find the probability that a sample of eight atoms was not in the ground state to be 4$\cdot 10^{-3}$. From spilling twice we also find an upper bound of $6$($2$)\% for the number of excitations during the ramps.

\subsection*{Imbalanced samples.}
We create imbalanced samples with a different number of atoms in states $\vert1\rangle$ and $\vert2\rangle$ in a two-step process. We first prepare a balanced system and subsequently remove additional atoms in state $\vert1\rangle$ by performing a second spilling process at a magnetic field of 40\,Gauss where the magnetic moment of state $\vert2\rangle$ vanishes (fig. S2). To estimate the preparation fidelity we consider the two components separately: The fidelity of the majority component ($\vert2\rangle$) is only given by the first spilling process since it is unaffected by the second spilling process. The uncertainty in the minority component ($\vert1\rangle$) is solely determined by the second spilling process. Thus the overall fidelity for the creation of an imbalanced system is similar to the balanced case.

\clearpage
\begin{figure}[htb!]
\centering
\includegraphics*[width=79mm]{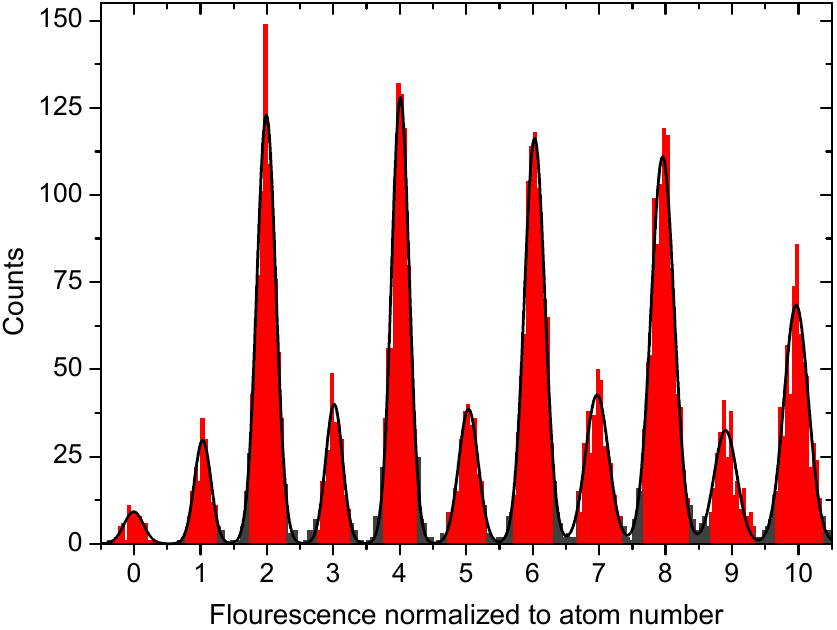}
\end{figure}
Figure S1:
Histogram of the rescaled fluorescence signals from the measurements shown in figure 2. The black curves are Gaussian fits to the rescaled atom number. The 2-atom (8-atom) peak center is separated from its adjacent peak centers by $7 \sigma$ ($5.7 \sigma$). This is large enough to clearly distinguish the fluorescence signal of different atom numbers. We bin fluorescence data within a $2 \sigma$-width of the peaks (red bars) to integer atom numbers; the counts outside the $2 \sigma$ widths(grey bars) are rejected.
\newpage
\begin{figure}[htb!]
\centering
\includegraphics*[width=79mm]{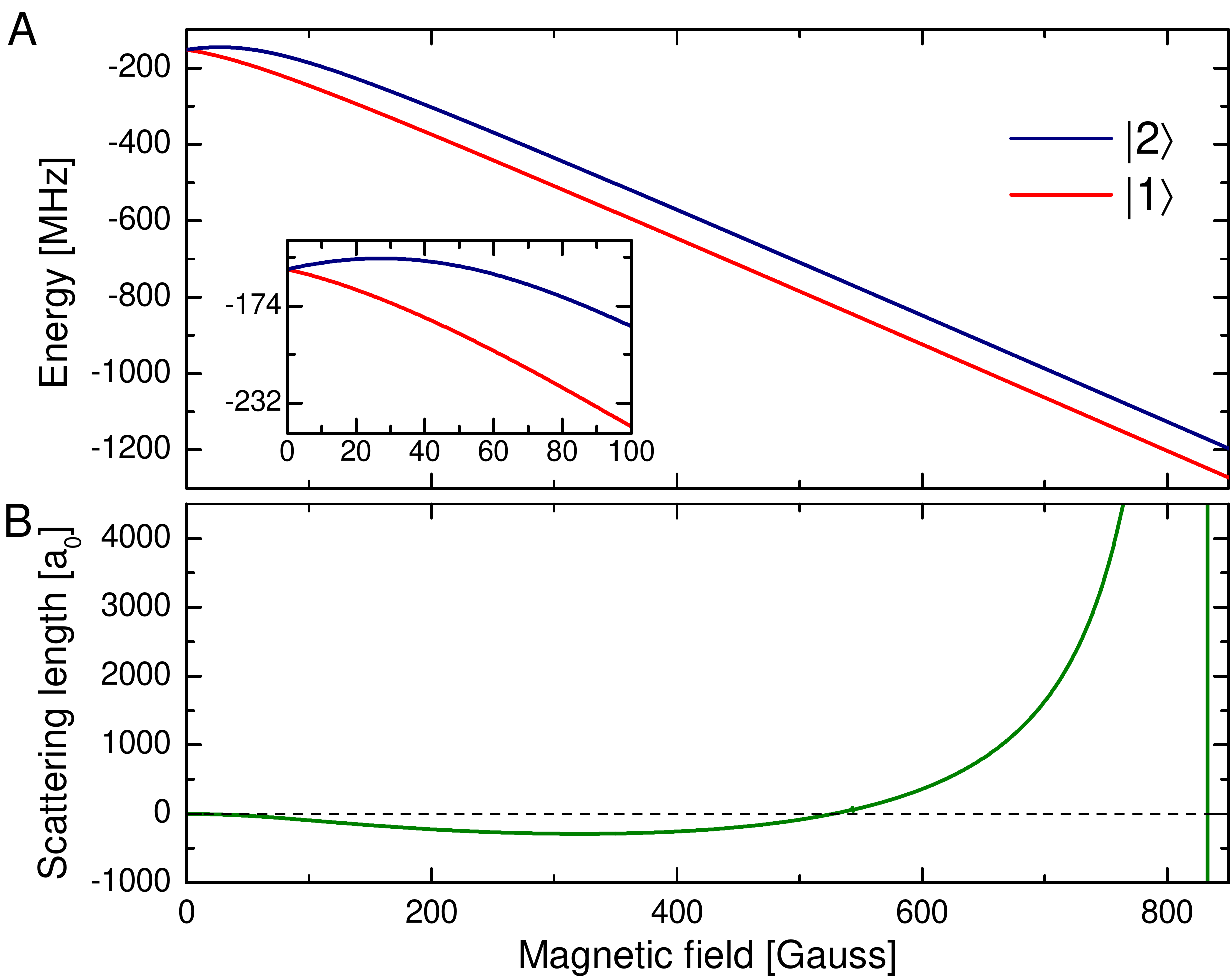}
\end{figure}
Figure S2:
Magnetic field dependence of the Zeeman sublevels and the scattering length. (A) Energy of the Zeeman sublevels of $^6$Li in the F=$1/2$ electronic ground state. The inset shows the magnetic field region below $100$\,G where the magnetic moment of state $\vert 2 \rangle$ crosses zero. (B) Two-body scattering length between atoms in states $\vert 1 \rangle$  and $\vert 2 \rangle$ as a function of the magnetic field. For evaporative cooling of the reservoir and for the transfer of atoms into the microtrap the offset field is tuned to $300$\,G. The spilling process is performed at a magnetic field of $523$\,G where the interaction strength vanishes because of the zero crossing of the scattering length. For the observation of the interaction-induced dynamics we tune to the magnetic field $760$\,G to obtain a pair of repulsively interacting atoms.

% For your review copy (i.e., the file you initially send in for
% evaluation), you can use the {figure} environment and the
% \includegraphics command to stream your figures into the text, placing
% all figures at the end.  For the final, revised manuscript for
% acceptance and production, however, PostScript or other graphics
% should not be streamed into your compliled file.  Instead, set
% captions as simple paragraphs (with a \noindent tag), setting them
% off from the rest of the text with a \clearpage as shown  below, and
% submit figures as separate files according to the Art Department's
% instructions.

\end{document}